\documentclass[showpacs,amssymb,preprint,preprintnumbers]{revtex4}

\usepackage{amsmath}

\begin{document}

\title{Graviton noise on tidal forces and geodesic congruences}
\author{Hing-Tong Cho}
\email[Email: ]{htcho@mail.tku.edu.tw}
\affiliation{Department of Physics, Tamkang University, Tamsui, New Taipei City, TAIWAN}
\author{Bei-Lok Hu}
\email[Email: ]{blhu@umd.edu}
\affiliation{Maryland Center for Fundamental Physics and Joint Quantum Institute, University of Maryland, College Park, Maryland 20742-4111, USA}

\begin{abstract}
In this work we continue with our recent study, using the Feynman-Vernon worldline influence action and the Schwinger-Keldysh closed-time-path formalism, to consider the effects of quantum noise of gravitons on the motion of point masses. This effect can be regarded as due to a stochastic tensorial force  whose correlator is given by the graviton noise kernel associated with the Hadamard function of the quantized gravitational field.  Solving the Langevin equation governing the motion of the separation of two masses, the fluctuations of the separation due to the graviton noise can be obtained for various states of the quantum  field. Since this force has the stretching and compressing effects like the tidal force, we can view it as one.  We therefore derive the expressions for, and estimate the magnitude of,  this tidal force for the cases of the Minkowski and the squeezed vacua. The  influence of this force on the evolution of the geodesic congruence through the Raychaudhuri equation  is then studied and the effects of quantum graviton noise on the shear and rotation tensors presented.
\end{abstract}

\date{January 16, 2023}
\maketitle

\section{Introduction}

Demonstrating the existence of gravitons is a necessity to affirming the quantum nature of {\it perturbative} gravity \cite{Feynman,Weinberg}.  This is in principle doable at today's low energy, unlike quantum gravity proper, such as string or loop quantum gravity, operative at the Planck scale \cite{Oriti,QGpheno,QGexpt}. Bounds on graviton's mass are deduced from  gravitational waves detected in LIGO \cite{gravitonGW}, but graviton in the classical context of gravitational waves invokes the assumption of a propagating particle which mediates gravitational forces, not in the full context of (perturbative) quantum gravity. Theoretically one can ask how graviton as a quantum entity changes the Newtonian potential,  as in, e.g., \cite{Don,HamLiu,DalMazNewton}; one can compute the graviton-induced corrections to the classical wave equations, as in, e.g., \cite{DalMazCarmen} for the Maxwell equation under the one-loop and weak field approximations. The corrected equations are reported to be analogous to the classical equations in anisotropic and inhomogeneous media. From the changes in the dispersion relations one can in principle induce the existence of gravitons.  However, direct experimental detection of gravitons remains a serious challenge \cite{Dyson}. 

Lately,  there is a surge of interest in accessible albeit indirect experimental detection of gravitons (see, e.g., \cite{QGTable,pertQG}). One proposal suggested by Parikh, Wilczek, and Zahariade \cite{PWZ20,PWZ21a,PWZ21b} is to measure the effects of graviton noise on  masses. This line of investigation was pursued by several groups of authors \cite{Kanno,Haba,ChoHu22} and will be the main focus of this  work.  In a recent paper  by the present  authors \cite{ChoHu22}, the effects of graviton noise on free test masses in Minkowski spacetime were considered via the Feynman-Vernon \cite{FeyVer} worldline influence functional and the Schwinger-Keldysh \cite{Schwinger,Keldysh} closed time path formalism \cite{Chou,CalHu08,HuVer20}  applied to the quantum Brownian motion model \cite{Schwinger,CalLeg,HPZ}. The noise kernel originating from the Hadamard function of the gravitons acts as a stochastic tensorial force in a Langevin equation governing the motion of the separation of two masses. The fluctuations of the separation due to the graviton noise are then solved for various quantum states including the Minkowski vacuum, thermal, coherent and squeezed states.

The purpose of this work is to study the effects of  graviton noise on the tidal
forces as well as the geodesic congruences. For this purpose we introduce a Newtonian potential $\phi(x)$ in the background.  The tidal force between two masses each following its geodesics arises from the  difference of forces exerted on the two masses, as governed by the geodesic deviation equation. Fluctuations of a quantum field, here the quantized linear perturbations of a weak gravitational field, constitute graviton noise.  At the lowest consistent order  we can neglect the interaction of the Newtonian potential with the gravitons. Thus we can derive graviton noise using  a Minkowski spacetime background and calculate its effect on the tidal force generated by the Newtonian potential. 

Following this,  we continue to investigate the effect of graviton noise on many masses each following a geodesic, namely, a geodesic congruence. The kinematics of a geodesic congruence, characterized by three parameters: an expansion scalar, a  shear tensor and a rotation tensor, collectively called the `deformation tensor', is governed by the Raychaudhuri equation \cite{RayEq} (for useful reviews see, e.g., \cite{Ellis,Kar}). The Raychaudhuri equation played an important role in the landmark proofs by Penrose, Hawking and Geroch of the existence of singularities in classical general relativity \cite{HawEll,RaySing}. Quantum versions have also been used to test out different quantum gravitational theories from `improved' (asymptotic safety) to `zero-point length' to generalized uncertainty principles and loop quantum gravity \cite{QRayEqDas,RayEqZPL,RayEqBGR,QRayEqGUP,RayEqEffGUP,RayEqLQG,RayEqEff}, to name just a few veins \footnote{Again, beware the fundamental differences between nonperturbative quantum gravity at the Planck scale  and low energy quantum perturbative gravity. Only the latter class is considered here. Also, the theoretical framework we use is quantum fields and their fluctuations (noise) in a fixed curved spacetime,  at the test-field level. The next two levels up including the backreaction of the mean value and the fluctuations of the stress energy tensor of quantum fields on the spacetime, at the semiclassical and stochastic gravity \cite{HuVer20} levels respectively, are not considered here. Fluctuations at the test field level are called intrinsic or active, fluctuations in the geometry at the stochastic gravity level are called induced or passive. In this context the Raychaudhuri equation will assume the form of a Langevin equation (see, e.g., \cite{Ford}).}.  Our goal is to find out how graviton noises  impact  on the deformation tensor components in both the Minkowski and the squeezed vacuum states. These are observable  quantities in physical systems at low energies which,  once  measured,  could reveal not only the existence, but also the nature and dynamics of gravitons through the action of their noises.

In Sec.~II, we review the the Feynman-Vernon influence functional formalism  implemented in \cite{ChoHu22} to derive the quantum noise effects of gravitons. It is shown that the influence of this noise to the geodesic separation of two masses is manifested as a stochastic tensor force in the Langevin equation of motion. In Sec.~III, we point out that the stochastic force can also be viewed as a tidal effect on neighboring masses. We then estimate the possibility of measuring this tidal force coming from gravitons in different quantum states of the field. It is natural to extend the consideration of neighboring particles to that of a geodesic congruence. This extension is carried out in Sec.~IV, where the evolution of the expansion scalar, the shear tensor as well as the rotation tensor are investigated through the Raychaudhuri equation with external forces. We conclude with   discussions in Sec.~V. \footnote{After we have completed the write-up of all the results reported here, while in the process of composing the Introduction and the Conclusion  we became aware of this new paper \cite{Bak} treating the same problem. We welcome this and hope it adds to the benefit of the readers.}

\section{Graviton noise}

To make our presentation self-contained we briefly describe the origin of graviton noises  and summarize their effects on free masses via the Langevin equation according to \cite{ChoHu22}. we begin with the Einstein action
\begin{eqnarray}
S_{g}=\frac{1}{\kappa^{2}}\int d^{4}x \sqrt{-g}\,R
\end{eqnarray}
where $\kappa^{2}=16\pi G$ and $G$ is the Newton's constant. Quantizing the linear gravitational perturbations, we obtain the action for the graviton field $h^{(s)}(x)$,
\begin{eqnarray}
S_{grav}=-\frac{1}{2}\int\,d^{4}x\,\sum_{s}\partial_{\alpha}h^{(s)}(x)\partial^{\alpha}h^{(s)}(x)\label{gaction}
\end{eqnarray}
which represents two minimally coupled massless scalar fields $h^{(s)}(x)$ for the two polarizations $s$ of the graviton.

In the Fermi normal coordinate $(t,\vec{z})$, the action of the test mass $m$ can be written as
\begin{eqnarray}
S_{m}&=&-m\int\,\sqrt{-ds^{2}}\nonumber\\
&=&\int\,dt\,\left[\frac{m}{2}\,\delta_{ij}\,\dot{ z}^{i}\dot{ z}^{j}+\frac{m\kappa}{4}\,\ddot{h}_{ij}\, z^{i} z^{j}\right]+\cdots\label{maction}
\end{eqnarray}
The first term is the kinetic term and the second one is the interaction term with the gravitational perturbation $h_{ij}(x)$. Using the scalar fields $h^{(s)}$, this interaction can be expressed as
\begin{eqnarray}
\int\,dt\,\frac{m\kappa}{4}\,\ddot{h}_{ij}\, z^{i} z^{j}=\alpha\int\,d^{4}x\,\sum_{s}h^{(s)}(x)X^{(s)}(x)\label{intterm}
\end{eqnarray}
where the constant $\alpha=m\kappa/2\sqrt{2}(2\pi)^3$ and 
\begin{eqnarray}
X^{(s)}(x)=\int\,d^{3}k\,e^{i\vec{k}\cdot\vec{x}}\frac{d^{2}}{dt^{2}}\left({\epsilon^{(s)}_{i}}^{*}(\vec{k}) z^{i}(t)\right)^{2}\label{defX}
\end{eqnarray}
with $\epsilon_{i}^{(s)}$ being the polarization vector corresponding to the graviton.

To examine the influence of the gravitons on the equation of motion of the test mass, we shall follow the in-in closed time path integral formalism. The interaction term in Eq.~(\ref{intterm}) is linear in the graviton field. Therefore, the graviton fields can be integrated over to produce the influence action
\begin{eqnarray}
S_{IF}&=&\int dt\,dt'\,\Delta^{ij}(t)D_{ijkl}(t,t')\Sigma^{kl}(t')+\frac{i}{2}\int dt\, dt'\,\Delta^{ij}(t)N_{ijkl}(t,t')\Delta^{kl}(t')\label{IFaction2}
\end{eqnarray}
where 
\begin{eqnarray}
\Sigma^{ij}(t)&=&\frac{1}{2}\left[ z_{+}^{i}(t) z_{+}^{j}(t)+ z_{-}^{i}(t) z_{-}^{j}(t)\right]\\\Delta^{ij}(t)&=& z_{+}^{i}(t) z_{+}^{j}(t)- z_{-}^{i}(t) z_{-}^{j}(t),
\end{eqnarray}
$ z_{+}^{i}(t)$ and $ z_{-}^{i}(t)$ are the forward and backward in-time fields in the closed-time path formalism. Here, 
\begin{eqnarray}
D_{ijkl}(t,t')=\alpha^{2}\frac{d^{2}}{dt^{2}}\frac{d^{2}}{dt'^{2}}\int d^{3}k\,d^{3}k'\int d^{3}x\,d^{3}x'\,e^{-i\vec{k}\cdot\vec{x}}e^{-i\vec{k}'\cdot\vec{x}'}\sum_{s}\epsilon_{ij}^{(s)}(\vec{k})\epsilon_{kl}^{(s)}(\vec{k}')G_{ret}(x,x')\nonumber\\
\end{eqnarray}
is the dissipation kernel with 
\begin{eqnarray}
G_{ret}(x,x')=i\theta(t-t')\langle[h(x),h(x')]\rangle
\end{eqnarray}
being the retarded Green function. 
\begin{eqnarray}
N_{ijkl}(t,t')=\frac{\alpha^{2}}{2}\frac{d^{2}}{dt^{2}}\frac{d^{2}}{dt'^{2}}\int d^{3}k\,d^{3}k'\int d^{3}x\,d^{3}x'\,e^{-i\vec{k}\cdot\vec{x}}e^{-i\vec{k}'\cdot\vec{x}'}\sum_{s}\epsilon_{ij}^{(s)}(\vec{k})\epsilon_{kl}^{(s)}(\vec{k}')G^{(1)}(x,x')\nonumber\\
\label{noise}
\end{eqnarray}
is the noise kernel where 
\begin{eqnarray}
G^{(1)}(x,x')=\langle\{h(x),h(x')\}\rangle\label{hadamard}
\end{eqnarray}
being the Hadamard function. Using the Feynman-Vernon Gaussian functional identity, the noise term can be replaced by a path integral over the stochastic tensor force $\xi_{ij}(t)$ with the two-point correlation function
\begin{eqnarray}
\langle\xi_{ij}(t)\xi_{kl}(t')\rangle_{s}
&=&N_{ijkl}(t,t')
\end{eqnarray}
Together with the kinetic terms of the test mass, we finally arrive at the so-called stochastic effective action $S_{SEA}$,
\begin{eqnarray}
S_{SEA}&=&S_{m}[ z_{+}]-S_{m}[ z_{-}]+S_{IF}\nonumber\\
&=&\frac{m}{2}\int dt\,\delta_{ij}\dot{ z}_{+}^{i}(t)\dot{ z}_{+}^{j}(t)-\frac{m}{2}\int dt\,\delta_{ij}\dot{ z}_{-}^{i}(t)\dot{ z}_{-}^{j}(t)\nonumber\\
&&\hskip 10pt + \int dt\,dt'\,\Delta^{ij}(t)D_{ijkl}(t,t')\Sigma^{kl}(t')-\int dt\,\xi_{ij}(t)\Delta^{ij}(t)
\end{eqnarray}
from which one can derive the effective equation of motion.

Indeed, the Langevin equation of motion for the test mass can be obtained by
\begin{eqnarray}
&&\left.\frac{\delta S_{SEA}}{\delta z^{i}_{+}}\right|_{ z_{+}= z_{-}= z}=0\nonumber\\
&\Rightarrow& m\ddot{ z}^{i}(t)+2\delta^{im}\int dt'\,D_{mnkl}(t,t')\, z^{n}(t) z^{k}(t') z^{l}(t')-2\delta^{ik}\xi_{kl}(t)\, z^{l}(t)=0\label{langevin1}
\end{eqnarray}
In the following we shall concentrate on the stochastic tensor force term coming from the effect of the graviton noise. That is, we shall neglect the history dependent dissiptation term which is supposed to be of higher order effect. Then, Eq.~(\ref{langevin1}) simplifies to
\begin{eqnarray}
m\ddot{ z}^{i}(t)= 2\delta^{ik}\xi_{kl}(t)\, z^{l}(t)\label{langevin2}
\end{eqnarray}

We shall explore in the subsequent sections the effects of the graviton noise on the tidal forces as well as the geodesic congruences. For this purpose we introduce a Newtonian potential $\phi(x)$ in the background and at the lowest consistent order we can neglect the interaction of the Newtonian potential with the graviton. In other words, gravitons come from the quantized linear perturbations of flat space, not  of the weakly curved space due to the gravitational potential, while  the gravitational potential remains a classical external field which mediates the tidal forces between two or more masses.  We want to know how gravitons influence the tidal forces between two masses and affect the congruences of masses in their geodesic motions, as a way to identify the presence of hitherto evasive gravitons. Therefore, we shall keep on using various vacuum states in the Minkowski spacetime for the evaluations of the Hadamard function and the noise kernel in Eqs.~(\ref{noise}) and (\ref{hadamard}).

With the Newtonian potential, the equation of motion is modified to 
\begin{eqnarray}
m\ddot{ z}^{i}(t)=-m\,\partial^{i}\phi(\vec{z})+ 2\delta^{ik}\xi_{kl}(t)\, z^{l}(t)\label{langevin3}
\end{eqnarray}
This is the form of the Langevin equation of motion of the test mass we shall consider in the following sections.

\section {Tidal forces}

In this section, we shall examine the effect of the graviton noise on the tidal forces experienced by test particles. Since tidal forces measure the difference in forces exerted to a material body at neighboring locations \cite{MTW}, we take the variation of $z^{i}(t)$ at fixed time $t$ of the modified Langevin equation of motion as shown in Eq.~(\ref{langevin3}).

\begin{eqnarray}
\frac{d^{2}}{dt^{2}}{(\Delta z)}^{i}&=&\left[-\partial^{i}\partial_{j}\phi+\left(\frac{2}{m}\right){\xi^{i}}_{l}\right](\Delta z)^{j}\nonumber\\
&\equiv&{K^{i}}_{j}(\Delta z)^{j}\label{tidaleqn}
\end{eqnarray}
where $(\Delta z)^{i}$ represents the geodesic deviation of a nearby geodesic in the congruence at time $t$ in the spatial directions. $-\partial_{i}\partial_{j}\phi$ is the Newtonian tidal tensor which is symmetric and traceless. Hence, the stochastic tensor force $2\xi_{ij}/m$ can also be interpreted as a form of tidal force. It is also symmetric and traceless. As a result the total tidal tensor $K_{ij}$ therefore consists of a classical Newtonian part and a stochastic part due to the gravitons.

The stochastic average of the tidal tensor
\begin{eqnarray}
\langle K_{ij}\rangle=-\partial_{i}\partial_{j}\phi
\end{eqnarray}
because the distribution of $\xi_{ij}$ is assumed to be Gaussian and $\langle\xi_{ij}\rangle=0$. The correlator of the stochastic part of the tidal tensor is just 
\begin{eqnarray}
\langle (\frac{2}{m}\xi_{ij}(t))(\frac{2}{m}\xi_{kl}(t')\rangle=\left(\frac{4}{m^{2}}\right)N_{ijkl}(t,t'),\label{stotidal}
\end{eqnarray}
given in terms of the noise kernel.

To have some estimation of the magnitudes of the tidal forces, we look at the typical example of that of a central mass $M$. The corresponding Newtonian potential would just be $\phi(r)=-\kappa^{2} M/16\pi r$. Hence, the classical Newtonian tidal tensor is given by
\begin{eqnarray}
-\partial_{i}\partial_{j}\phi(r)=-\frac{\kappa^{2} M}{16\pi r^3}\left(\delta_{ij}-\frac{3x_{i}x_{j}}{r^{2}}\right)
\end{eqnarray}
For radial geodesics, we take $x_{i}=(0,0,r)$, and 
\begin{eqnarray}
-\partial_{i}\partial_{j}\phi(r)=-\frac{\kappa^{2} M}{16\pi r^{3}}\left(
\begin{array}{ccc}
1 & 0 & 0 \\
0 & 1 & 0 \\
0 & 0 & -2
\end{array}
\right)
\end{eqnarray}
The tidal stress in the radial direction is positive, $\kappa^{2} M/8\pi r^3$, which represents stretching tension. In the transverse directions, the tidal stress is negative, $-\kappa^{2} M/16\pi r^3$, representing a compression strain.

One can therefore estimate the magnitude of this tidal tensor component by $\kappa^{2}M/16\pi r^3$. With $\kappa^{2}/16\pi\approx 6.67\times 10^{-11}$N m$^{2}\,$kg$^{-2}$ being the gravitational constant and $M_{\odot}\approx 2\times 10^{30}$~kg the mass of the sun,
\begin{eqnarray}
\frac{\kappa^{2}M}{16\pi r^{3}}\approx 1.33\times 10^{11}\left(\frac{M/M_{\odot}}{(r/1\, {\rm km})^{3}}\right)\ {\rm N}\,{\rm m}^{-1}{\rm kg}^{-1}
\end{eqnarray}
For example, near the event horizon ($r_{H}=3$ km) of a solar mass black hole,
\begin{eqnarray}
\frac{\kappa^{2}M}{16\pi r^{3}}\approx 4.93\times 10^{8}\ {\rm N}\,{\rm m}^{-1}{\rm kg}^{-1}\label{tidhor}
\end{eqnarray}
Therefore, for an object of mass $m=100$ kg and size $l=1$ m, the pressure due to the tidal stress near the horizon of a solar mass black hole, is given by
\begin{eqnarray}
\frac{\kappa^{2}Mm\,l}{16\pi r^{3}l^{2}}
\approx 4.93\times 10^{4}\ {\rm atm}
\end{eqnarray}
In fact, a human body cannot withstand a pressure of more than about 100 atmospheric pressures without breaking. This corresponds to a distance of about 100 km from the center of the solar mass black hole. 

To have another order of magnitude estimation, consider one near the surface of the earth with $M\approx M_{\odot}/333000$ and $r\approx 6400$ km,
\begin{eqnarray}
\frac{\kappa^2 M}{16\pi r^{3}}&\approx &1.33\times 10^{11}\left(\frac{1/333000}{(6400)^{3}}\right)\ {\rm N}\,{\rm m}^{-1}{\rm kg}^{-1}\nonumber\\
&\approx &1.52\times 10^{-6}\ {\rm N}\,{\rm m}^{-1}{\rm kg}^{-1}.\label{earthsurf}
\end{eqnarray}
In this case, the result is about 14 order of magnitude smaller than that in Eq.~(\ref{tidhor}).

Next, we would like to estimate the tidal forces due to the graviton noise as indicated in the correlator in Eq.~(\ref{stotidal}). In \cite{ChoHu22}, we have calculated the noise kernel from gravitons in the Minkowski vacuum,
\begin{eqnarray}
N_{ijkl}^{(0)}(t,t')=-\left(\frac{32\pi^{4}}{15}\right)\alpha^{2}\Lambda^{6}\,[2\,\delta_{ij}\delta_{kl}-3(\delta_{ik}\delta_{jl}+\delta_{il}\delta_{jk})]\,F[\Lambda(t-t')],\label{vacnoi}
\end{eqnarray}
where 
\begin{eqnarray}
F(x)&=&\frac{1}{x^{6}}\int_{0}^{x}dy\,y^{5}\cos y\nonumber\\
&=&\frac{1}{x^{6}}\left[(5x^{4}-60x^{2}+120)\cos x+x(x^{4}-20x^{2}+120)\sin x-120\right],\label{defF}
\end{eqnarray}
and $\Lambda$ is a momentum cutoff which will be related to some scale in the problem. To make the estimation, we consider the typical component $3333$ of the noise kernel in the coincident limit,
\begin{eqnarray}
N_{3333}^{(0)}(t,t)=\left(\frac{64\pi^{4}}{45}\right)\alpha^{2}\Lambda^{6},
\end{eqnarray}
which is independent of the time $t$ because the Minkowski vacuum is time-translation invariant. 

Now, we estimate the magnitude of the stochastic part of the tidal tensor component by taking the square root of the fluctuation, that is, the coincident limit of the correlation function. Remember the constant $\alpha=m\kappa/2\sqrt{2}(2\pi)^{3}$ so
\begin{eqnarray}
\sqrt{\hbar c\left(\frac{4}{m^{2}}N_{3333}^{(0)}(t,t)\right)}=\sqrt{\frac{\hbar c \kappa^2\Lambda^{6}}{90\pi^{2}}},
\end{eqnarray}
where we have added the factor $\hbar c$ to obtain the right dimension. If we take the momentum cutoff as the reciprocal of the typical length in the problem, that is, $\Lambda\sim 1/l\sim 1$ m$^{-1}$, then
\begin{eqnarray}
\sqrt{\hbar c\left(\frac{4}{m^{2}}N_{3333}^{(0)}(t,t)\right)}=2.39\times 10^{-18}\ {\rm N}\,{\rm m}^{-1}{\rm kg}^{-1}\label{tidgra}
\end{eqnarray}
which is about 12 order of magnitude smaller than that near the Earth's surface as shown in Eq.~(\ref{earthsurf}). 

In \cite{ChoHu22}, we have also discussed the case of squeezed vacuum state. The explicit form of the corresponding noise kernel is 
\begin{eqnarray}
N_{ijkl}^{( \zeta )}(t,t')
&=&(\cosh2 \zeta )N_{ijkl}^{(0)}(t,t')\nonumber\\
&&\ \ -(\sinh2 \zeta )B_{ijkl}\left(\frac{16\pi^{4}}{15}\right)\alpha^{2}\Lambda^{6}F[\Lambda(t+t')].\label{squeezedNK}
\end{eqnarray}
where $\zeta$ is the squeeze parameter and the tensor $B_{ijkl}$ is
\begin{eqnarray}
B_{ijkl}&=&(\delta_{ij}\delta_{kl}+\delta_{ik}\delta_{jl}+\delta_{il}\delta_{jk})-5\,[\delta_{ij}(\hat{ u}_{0})_{k}(\hat{ u}_{0})_{l}+\delta_{ik}(\hat{ u}_{0})_{j}(\hat{ u}_{0})_{l}\nonumber\\
&&\ \ \ \ \ +\delta_{il}(\hat{ u}_{0})_{j}(\hat{ u}_{0})_{k}+\delta_{jk}(\hat{ u}_{0})_{i}(\hat{ u}_{0})_{l}+\delta_{jl}(\hat{ u}_{0})_{i}(\hat{ u}_{0})_{k}+\delta_{kl}(\hat{ u}_{0})_{i}(\hat{ u}_{0})_{j}]\nonumber\\
&&\ \ \ \ \ +35\,(\hat{ u}_{0})_{i}(\hat{ u}_{0})_{j}(\hat{ u}_{0})_{k}(\hat{ u}_{0})_{l},\label{defB}
\end{eqnarray}
with $\hat{u}_{0}$ being an arbitrary unit vector. It is usually chosen to be $(0,0,1)$. Again, we examine the component $3333$ of the noise kernel in the coincident limit. Note that the tensor in the squeezed state noise kernel $B_{3333}=8$. Then,
\begin{eqnarray}
N_{3333}^{(\zeta)}(t,t)=(\cosh2\zeta)\left(\frac{64\pi^{4}}{45}\right)\alpha^{2}\Lambda^{6}-(\sinh 2\zeta)\left(\frac{128\pi^{4}}{15}\right)\alpha^{2}\Lambda^{6}F(2\Lambda t).
\end{eqnarray}
Note that here the coincident limit of the noise kernel is no longer independent of the time $t$ since the squeezed vacuum state is not time-tranlation invariant.
Suppose we take the typical length and time scale both to be $l,t\sim 1/\Lambda\sim 1$ m. The function
\begin{eqnarray}
F(2\Lambda t)&\approx&-\frac{1}{8}\left[5\cos(2)-14\sin(2)+15\right]\nonumber\\
&\approx&-0.0236
\end{eqnarray}

We can see that the squeezed state noise kernel is basically enhanced by the exponential factor $e^{2\zeta}$ as compared to that in the Minkowski vacuum. Hence, the magnitude of the stochastic part of the tidal tensor component can be similarly enhanced by $e^{\zeta}$. In many inflationary cosmological scenarios, the primordial gravitons produced in the early universe are in a squeezed state with a large squeeze parameter. Hence, this enhancement due to squeezing might be enough to augment the graviton quantum noise induced tidal force to a detectable level.

\section{Geodesic congruences}

In the last section we have discussed the effect of quantum noise due to gravitons on the motion of test particles manifesting itself in the form of tidal forces. The presence of these tidal forces would in turn influence the evolution of the geodesic congruences. Hence, in this section we would consider further the so-called deformation tensor ${\Omega^{i}}_{j}$ which characterizes the shape of the congruence.

One can define the deformation tensor ${\Omega^{i}}_{j}$ by the equation
\begin{eqnarray}
\frac{d(\Delta z)^{i}}{dt}={\Omega^{i}}_{j}(\Delta z)^{j}\label{defOmega}
\end{eqnarray}
Then the relationship between $\Omega_{ij}$ and the tidal tensor $K_{ij}$ can be established as follows \cite{Shore18}. Differentiate Eq.~(\ref{defOmega}) with respect to $t$ and we have
\begin{eqnarray}
\frac{d^{2}(\Delta z)^{i}}{dt^{2}}=\left(\frac{d{\Omega^{i}}_{j}}{dt}+{\Omega^{i}}_{k}{\Omega^{k}}_{j}\right)(\Delta z)^{j}.
\end{eqnarray}
Comparing this with the tidal equation in Eq.~(\ref{tidaleqn}), the evolution of the deformation tensor can be expressed as
\begin{eqnarray}
\frac{d{\Omega^{i}}_{j}}{dt}=-{\Omega^{i}}_{k}{\Omega^{k}}_{j}+{K^{i}}_{j}\label{evoleqn}
\end{eqnarray}

Usually, $\Omega_{ij}$ is decomposed into its trace, symmetric-traceless, and antisymmetric parts.
\begin{eqnarray}
\theta&=&{\Omega^{i}}_{i}\nonumber\\
\sigma_{ij}&=&\frac{1}{2}\left(\Omega_{ij}+\Omega_{ji}\right)-\frac{1}{2}{\Omega^{k}}_{k}\delta_{ij}\nonumber\\
\omega_{ij}&=&\frac{1}{2}\left(\Omega_{ij}-\Omega_{ji}\right)
\end{eqnarray}
with $\theta$ being the expansion scalar, $\sigma_{ij}$ the shear tensor, and $\omega_{ij}$ the rotation tensor \cite{Poisson}. That is, 
\begin{eqnarray}
\Omega_{ij}=\frac{1}{3}\theta\delta_{ij}+\sigma_{ij}+\omega_{ij}.
\end{eqnarray}

The evolution equations for various parts listed above can be obtained from Eq.~(\ref{evoleqn}). 
For the expansion scalar, which measures the rate of change in the volume of the congruence,
\begin{eqnarray}
\dot{\theta}&=&-{\Omega^{i}}_{j}{\Omega^{j}}_{i}+{K^{i}}_{i}\nonumber\\
&=&-\frac{1}{3}\theta^{2}-\sigma_{ij}\sigma^{ij}+\omega_{ij}\omega^{ij}+{K^{i}}_{i}.\label{expevo}
\end{eqnarray}
This is just the Raychaudhuri equation with an external force.

For the shear tensor,
\begin{eqnarray}
\dot{\sigma}_{ij}&=&-\frac{1}{2}\Omega_{ik}{\Omega^{k}}_{j}+\frac{1}{2}K_{ij}-\frac{1}{6}\delta_{ij}\left(-\frac{1}{3}\theta^{2}-\sigma_{kl}\sigma^{kl}+\omega_{kl}\omega^{kl}+{K^{k}}_{k}\right)+(i\leftrightarrow j)\nonumber\\
&=&-\frac{2}{3}\theta\sigma_{ij}-\sigma_{ik}{\sigma^{k}}_{j}-\omega_{ik}{\omega^{k}}_{j}+\frac{1}{3}\delta_{ij}\left(\sigma_{kl}\sigma^{kl}-\omega_{kl}\omega^{kl}\right)
+\frac{1}{2}K_{ij}+\frac{1}{2}K_{ji}-\frac{1}{3}{K_{k}}^{k}\delta_{ij}\nonumber\\
\label{sheevo}
\end{eqnarray}

For the rotation tensor,
\begin{eqnarray}
\dot{\omega}_{ij}&=&\frac{1}{2}\left(-\Omega_{ik}{\Omega^{k}}_{j}+K_{ij}\right)-(i\leftrightarrow j)\nonumber\\
&=&-\frac{2}{3}\theta\omega_{ij}-\sigma_{ik}{\omega^{k}}_{j}-\omega_{ik}{\sigma^{k}}_{j}+\left(K_{ij}+K_{ji}\right)\label{rotevo}
\end{eqnarray}

This set of equations shows how the evolution of the deformation tensor is affected by the tidal forces. Suppose that there is no Newtonian potential or that the Newtonian gravitational force is uniform, then we have $-\partial_{i}\partial_{j}\phi=0$. In this case, the tidal tensor is given solely by the stochastic force 
\begin{eqnarray}
K_{ij}=\left(\frac{2}{m}\right)\xi_{ij},
\end{eqnarray}
induced by the graviton noise. In the following subsections we shall concentrate on this case to investigate what kind of impact the graviton noise would have on the deformation tensor components in both the Minkowski and the squeezed vacuum states.

\subsection{The expansion scalar and the rotation tensor}
We start with the evolution equation of the expansion scalar in Eq.~(\ref{expevo}).
Suppose that the optical tensor components are small compared to the terms given by the tidal force. Then, one can solve this evolution equation perturbatively as
\begin{eqnarray}
\dot{\theta}\sim{K^{i}}_{i}\Rightarrow\theta=\int^{t}_{0}dt'{K^{i}}_{i}(t')+\theta_{0}
\end{eqnarray}
where $\theta_{0}$ is its initial value. Therefore, the influence of the gravitons on the expansion scalar can be expressed as
\begin{eqnarray}
\theta=\frac{2}{m}\int_{0}^{t}dt'\,{\xi_{i}}^{i}(t')+\theta_{0}
\end{eqnarray}

Under the stochastic average, we have
\begin{eqnarray}
\langle\theta\rangle&=&\theta_{0}\\
\langle\theta(t)\theta(t')\rangle&=&\theta_{0}^{2}+\left(\frac{2}{m}\right)^{2}\int_{0}^{t}dt''\int_{0}^{t'}dt'''\langle{\xi_{i}}^{i}(t''){\xi_{j}}^{j}(t''')\rangle\nonumber\\
&=&\theta_{0}^{2}+\frac{4}{m^{2}}\int_{0}^{t}dt''\int_{0}^{t'}dt'''\,{{{N_{i}}^{i}}_{j}}^{j}(t'',t''')
\end{eqnarray}
The correlation function of the expansion scalar is thus given by the integrations over the noise kernel.

The explicit form of the noise kernel $N^{(0)}_{ijkl}$ due to gravitons in the Minkowski vacuum is given in Eq.~(\ref{vacnoi}). From the symmetry structure of this noise kernel, we can see that it is symmetric and traceless with respect to the pairs of indices $ij$ and $kl$. This is because the noise kernel is related to the influence of the gravitons on the test particles, while the graviton is represented by a symmetric-traceless tensor field. Hence, these symmetries of the noise kernel is true not only for the vacuum state but also for any other quantum states of the corresponding graviton. From the traceless property, we can see that the noise kernel will not contribute to the correlator of the expansion scalar $\theta$. That is, 
\begin{eqnarray}
\langle\theta(t)\theta(t')\rangle^{(0)}&=&\theta_{0}^{2}
\end{eqnarray}

We can also consider the squeezed state. From the explicit expression for the squeezed state noise kernel in Eq.~(\ref{squeezedNK}), it is apparent that the tensor $B_{ijkl}$ is also symmetric and traceless with respect to the pairs $ij$ and $kl$. Therefore, the noise kernel in the squeezed state will not contribute to the correlators of the expansion scalar, and 
\begin{eqnarray}
\langle\theta(t)\theta(t')\rangle^{(\zeta)}&=&\theta_{0}^{2}
\end{eqnarray}

The same consideration can be applied to the rotation tensor. From the evolution equation in Eq.~(\ref{rotevo}) and concentrating on the effect of tidal force, we have
\begin{eqnarray}
\dot{\omega}_{ij}\sim\frac{1}{2}\left(K_{ij}+K_{ji}\right)\Rightarrow\omega_{ij}=\frac{1}{2}\int_{0}^{t}dt'\left(K_{ij}(t')+K_{ji}(t')\right)+({\omega_{0}})_{ij}
\end{eqnarray}
where $(\omega_{0})_{ij}$ is its initial value. Expressing the tidal force induced by the graviton noise with the stochastic force $\xi_{ij}$, 
\begin{eqnarray}
\omega_{ij}=\frac{1}{m}\int_{0}^{t}dt'\left(\xi_{ij}-\xi_{ji}\right)+(\omega_{0})_{ij},
\end{eqnarray}
with the stochastic averages
\begin{eqnarray}
\langle\omega_{ij}\rangle&=&(\omega_{0})_{ij}\\
\langle\omega_{ij}(t)\omega_{kl}(t')\rangle&=&(\omega_{0})_{ij}(\omega_{0})_{kl}+\frac{4}{m^{2}}\int_{0}^{t}dt''\int_{0}^{t'}dt'''\,N_{[ij][kl]}(t'',t''').
\end{eqnarray}
As we have discussed earlier, the noise kernels $N_{ijkl}^{(0)}$ and $N_{ijkl}^{(\zeta)}$, from graviton noise for the Minkowksi vacuum and the squeezed vacuum, respectively, are symmetric with respect to the pairs of indices $ij$ and $kl$. Hence, it will not contribute to the stochastic average above, and we have
\begin{eqnarray}
\langle\omega_{ij}(t)\omega_{kl}(t')\rangle^{(0)}=\langle\omega_{ij}(t)\omega_{kl}(t')\rangle^{(\zeta)}=(\omega_{0})_{ij}(\omega_{0})_{kl}
\end{eqnarray}

\subsection{The shear tensor}
Finally, we come to the shear tensor which is traceless and symmetric. Since the noise kernel due to graviton noise is also traceless and symmetric, we would expect the stochastic average of the shear tensor to have contributions from this noise. From the evolution equation in Eq.~(\ref{sheevo}), we have
\begin{eqnarray}
&&\dot{\sigma}_{ij}\sim\frac{1}{2}K_{ij}+\frac{1}{2}K_{ji}-\frac{1}{3}{K_{k}}^{k}\delta_{ij}\nonumber\\
&
\Rightarrow&\sigma_{ij}=\frac{1}{2}\int^{t}_{0}dt'\left(K_{ij}(t')+K_{ji}(t')-\frac{2}{3}\delta_{ij}{K_{k}}^{k}(t')\right)+({\sigma_{0}})_{ij}
\end{eqnarray}
where $({\sigma_{0}})_{ij}$ is its initial value. With the stochastic forces,
\begin{eqnarray}
\sigma_{ij}=\frac{1}{m}\int_{0}^{t}dt'\left[\xi_{ij}(t')+\xi_{ji}(t')-\frac{2}{3}\delta_{ij}{\xi_{k}}^{k}(t')\right]+(\sigma_{0})_{ij}
\end{eqnarray}
and the stochastic averages are
\begin{eqnarray}
\langle\sigma_{ij}\rangle&=&(\sigma_{0})_{ij}\\
\langle\sigma_{ij}(t)\sigma_{kl}(t')\rangle&=&(\sigma_{0})_{ij}(\sigma_{0})_{kl}+\frac{4}{m^{2}}\int_{0}^{t}dt''\int_{0}^{t'}dt'''\bigg[N_{(ij)(kl)}(t'',t''')
-\frac{1}{3}{N_{(ij)n}}^{n}(t'',t''')\delta_{kl}
\nonumber\\
&&\hskip 90pt-\frac{1}{3}{{N_{m}}^{m}}_{(kl)}(t'',t''')\delta_{ij}+\frac{1}{9}N_{m\hskip 9pt n}^{\hskip 7pt m\hskip 7pt n}(t'',t''')\delta_{ij}\delta_{kl}\bigg]\nonumber\\
&=&(\sigma_{0})_{ij}(\sigma_{0})_{kl}+\frac{4}{m^{2}}\int_{0}^{t}dt''\int_{0}^{t'}dt'''\,N_{ijkl}(t'',t''')
\label{shearcor}
\end{eqnarray}

First, we consider the correlator with gravitons in the Minkowski vacuum. 
Putting the expression for the noise kernel in Eq.~(\ref{vacnoi}) into the shear tensor correlator in Eq.~(\ref{shearcor}),  
\begin{eqnarray}
\langle\sigma_{ij}(t)\sigma_{kl}(t')\rangle^{(0)}&=&(\sigma_{0})_{ij}(\sigma_{0})_{kl}\nonumber\\
&&\ \ -\frac{128\pi^{4}\alpha^{2}\Lambda^{6}}{15m^{2}}\left[2\,\delta_{ij}\delta_{kl}-3(\delta_{ik}\delta_{jl}+\delta_{il}\delta_{jk})\right]\int_{0}^{t}dt''\int_{0}^{t'}dt'''\,F[\Lambda(t''-t''')]\nonumber\\
\end{eqnarray}
The integrations over $t''$ and $t'''$ can be done in closed form. However, if we expand the result in powers of $(t-t')$, we get
\begin{eqnarray}
&&\int_{0}^{t}dt''\int_{0}^{t'}dt'''\,F[\Lambda(t''-t''')]\nonumber\\
&=&\frac{1}{2\Lambda^{6}t^{4}}\left[\Lambda^{4}t^{4}-4\Lambda^{3}t^{3}\sin(\Lambda t)-12\Lambda^{2}t^{2}\cos(\Lambda t)+24\Lambda t\sin(\Lambda t)-24(1-\cos(\Lambda t))\right]\nonumber\\
&&\ \ +\frac{(t-t')}{\Lambda^{6}t^{5}}\left[\Lambda^{4}t^{4}\cos(\Lambda t)-4\Lambda^{3}t^{3}\sin(\Lambda t)-12\Lambda^{2}t^{2}\cos(\Lambda t)\right.\nonumber\\
&&\hskip 150pt\left.+24\Lambda t \sin(\Lambda t)-24(1-\cos(\Lambda t))\right]\nonumber\\
&&\ \ +\dots\label{firstint}
\end{eqnarray}
From this we can see that the shear tensor correlator $\langle\sigma_{ij}(t)\sigma_{kl}(t')\rangle^{(0)}$ in the Minkowski vacuum is finite in the coincident limit $t'\rightarrow t$. Hence,
\begin{eqnarray}
&&\langle\sigma_{ij}(t)\sigma_{kl}(t)\rangle^{(0)}\nonumber\\
&=&(\sigma_{0})_{ij}(\sigma_{0})_{kl}\nonumber\\
&&\ \ -\left[2\,\delta_{ij}\delta_{kl}-3(\delta_{ik}\delta_{jl}+\delta_{il}\delta_{jk})\right]\left(\frac{\kappa^{2}}{120\pi^{2}t^{4}}\right)\nonumber\\
&&\hskip 50pt\times\left[\Lambda^{4}t^{4}-4\Lambda^{3}t^{3}\sin(\Lambda t)-12\Lambda^{2}t^{2}\cos(\Lambda t)+24\Lambda t\sin(\Lambda t)-24(1-\cos(\Lambda t))\right]\nonumber\\
\end{eqnarray}
where we have substituted $\alpha=m\kappa/2\sqrt{2}(2\pi)^{3}$ to arrive at this expression. 

First, to concentrate on the stochastic part, we shall assume that $(\sigma_{0})_{ij}=0$. Then, the 3333 component of the correlator is given by
\begin{eqnarray}
&&\langle\sigma_{33}(t)\sigma_{33}(t)\rangle^{(0)}\nonumber\\
&=&\left(\frac{\kappa^{2}}{30\pi^{2}t^{4}}\right)\left[\Lambda^{4}t^{4}-4\Lambda^{3}t^{3}\sin(\Lambda t)-12\Lambda^{2}t^{2}\cos(\Lambda t)+24\Lambda t\sin(\Lambda t)-24(1-\cos(\Lambda t))\right]\nonumber\\
\end{eqnarray}
In fact, for the other components,
\begin{eqnarray}
&&\langle\sigma_{11}(t)\sigma_{11}(t)\rangle^{(0)}=\langle\sigma_{22}(t)\sigma_{22}(t)\rangle^{(0)}=\langle\sigma_{33}(t)\sigma_{33}(t)\rangle^{(0)}\nonumber\\
&&\langle\sigma_{11}(t)\sigma_{22}(t)\rangle^{(0)}=\langle\sigma_{22}(t)\sigma_{33}(t)\rangle^{(0)}=\langle\sigma_{33}(t)\sigma_{11}(t)\rangle^{(0)}=-\frac{1}{2}\langle\sigma_{33}(t)\sigma_{33}(t)\rangle^{(0)}\nonumber\\
&&\langle\sigma_{12}(t)\sigma_{12}(t)\rangle^{(0)}=\langle\sigma_{23}(t)\sigma_{23}(t)\rangle^{(0)}=\langle\sigma_{31}(t)\sigma_{31}(t)\rangle^{(0)}=\frac{3}{4}\langle\sigma_{33}(t)\sigma_{33}(t)\rangle^{(0)}
\end{eqnarray}
To estimate the magnitude of the shear, we take the momentum cutoff, as in our last section for the tidal force, to be $\Lambda\sim 1/l\sim 1$ m$^{-1}$. Furthermore, suppose that the time of duration of measurement $t$ to be also of the order of $l$, then the square root of the typical 3333 component
\begin{eqnarray}
\sqrt{\langle\sigma_{33}(t)\sigma_{33}(t)\rangle^{(0)}}&\sim&\sqrt{\left(\frac{\hbar}{c}\right)\frac{\kappa^{2}}{30\pi^{2}}}\nonumber\\
&\approx&1.94\times 10^{-27}\,{\rm s}^{-1}
\end{eqnarray}
where a factor of $\hbar/c$ has been added to have the right dimension. This estimation is consistent with that of the tidal force in Eq.~(\ref{tidgra}) as the momentum cutoff there is also taken to be $\sim$ 1 m$^{-1}$. Indeed, $(2.39\times 10^{-18}$ s$^{-2})(l/c)\sim 10^{-27}$ s$^{-1}$.

Lastly, we consider the shear tensor in the squeezed vacuum state. Concentrating on the stochastic part of the shear tensor correlator, that is, taking $(\sigma_{0})_{ij}=0$,
\begin{eqnarray}
&&\langle\sigma_{ij}(t)\sigma_{kl}(t')\rangle^{(\zeta)}\nonumber\\
&=&-(\cosh2\zeta)\left(\frac{128\pi^{4}\alpha^{2}\Lambda^{6}}{15m^{2}}\right)\left[2\,\delta_{ij}\delta_{kl}-3(\delta_{ik}\delta_{jl}+\delta_{il}\delta_{jk})\right]\int_{0}^{t}dt''\int_{0}^{t'}dt'''\,F[\Lambda(t''-t''')]\nonumber\\
&&\ \ -(\sinh2 \zeta )B_{ijkl}\left(\frac{64\pi^{4}\alpha^{2}\Lambda^{6}}{15m^{2}}\right)\int_{0}^{t}dt''\int_{0}^{t'}dt'''F[\Lambda(t''+t''')]
\end{eqnarray}
The first integral has been evaluated in Eq.~(\ref{firstint}). Similarly, we expand the second integral in powers of $t-t'$,
\begin{eqnarray}
&&\int_{0}^{t}dt''\int_{0}^{t'}dt'''\,F[\Lambda(t''+t''')]\nonumber\\
&=&-\frac{1}{8\Lambda^{6}t^{4}}[2\Lambda^{4}t^{4}-4\Lambda^{3}t^{3}(4\sin(\Lambda t)-\sin(2\Lambda t))-6\Lambda^{2}t^{2}(8\cos(\Lambda t)-\cos(2\Lambda t))\nonumber\\
&&\hskip 60pt +6\Lambda t\,(16\sin(\Lambda t)-\sin(2\Lambda t))-(93-96\cos(\Lambda t)+3\cos(2\Lambda t))]\nonumber\\
&&-\frac{(t-t')}{4\Lambda^{6}t^{5}}\,[2\Lambda^{4}t^{4}(2\cos(\Lambda t)-\cos(2\Lambda t))-4\Lambda^{3}t^{3}(4\sin(\Lambda t)-\sin(2\Lambda t))\nonumber\\
&&\hskip 60pt -6\Lambda^{2}t^{2}(8\cos(\Lambda t)-\cos(2\Lambda t))+6\Lambda t(16\sin(\Lambda t)-\sin(2\Lambda t))\nonumber\\
&&\hskip 60pt +(-93+96\cos(\Lambda t)-3\cos(2\Lambda t))]\nonumber\\
&&+\cdots
\end{eqnarray}

Therefore, in the coincident limit $t'\rightarrow t$,
\begin{eqnarray}
&&\langle\sigma_{ij}(t)\sigma_{kl}(t)\rangle^{(\zeta)}\nonumber\\
&=&-(\cosh2\zeta)\left[2\,\delta_{ij}\delta_{kl}-3(\delta_{ik}\delta_{jl}+\delta_{il}\delta_{jk})\right]\left(\frac{\kappa^{2}}{120\pi^{2}t^{4}}\right)\nonumber\\
&&\hskip 50pt\times\left[\Lambda^{4}t^{4}-4\Lambda^{3}t^{3}\sin(\Lambda t)-12\Lambda^{2}t^{2}\cos(\Lambda t)+24\Lambda t\sin(\Lambda t)-24(1-\cos(\Lambda t))\right]\nonumber\\
&&+(\sinh2 \zeta )B_{ijkl}\left(\frac{\kappa^{2}}{960\pi^{2}t^{4}}\right)[2\Lambda^{4}t^{4}-4\Lambda^{3}t^{3}(4\sin(\Lambda t)-\sin(2\Lambda t))\nonumber\\
&&\hskip 130pt -6\Lambda^{2}t^{2}(8\cos(\Lambda t)-\cos(2\Lambda t))+6\Lambda t\,(16\sin(\Lambda t)-\sin(2\Lambda t))\nonumber\\
&&\hskip 130pt -(93-96\cos(\Lambda t)+3\cos(2\Lambda t))]
\end{eqnarray}
To estimate the magnitude of the shear tensor fluctuation, we again take the 3333 component. 
\begin{eqnarray}
&&\langle\sigma_{33}(t)\sigma_{33}(t)\rangle^{(\zeta)}\nonumber\\
&=&(\cosh2\zeta)\left(\frac{\kappa^{2}}{30\pi^{2}t^{4}}\right)
[\Lambda^{4}t^{4}-4\Lambda^{3}t^{3}\sin(\Lambda t)-12\Lambda^{2}t^{2}\cos(\Lambda t)\nonumber\\
&&\hskip 150pt +24\Lambda t\sin(\Lambda t)-24(1-\cos(\Lambda t))
]\nonumber\\
&&+(\sinh2 \zeta )\left(\frac{\kappa^{2}}{120\pi^{2}t^{4}}\right)[2\Lambda^{4}t^{4}-4\Lambda^{3}t^{3}(4\sin(\Lambda t)-\sin(2\Lambda t))\nonumber\\
&&\hskip 130pt -6\Lambda^{2}t^{2}(8\cos(\Lambda t)-\cos(2\Lambda t))+6\Lambda t\,(16\sin(\Lambda t)-\sin(2\Lambda t))\nonumber\\
&&\hskip 130pt -(93-96\cos(\Lambda t)+3\cos(2\Lambda t))]
\end{eqnarray}
The first term is just the Minkowski vacuum result times $\cosh 2\zeta$. If we take the length and time scales to be $l\sim t\sim 1$ m and $\Lambda\sim 1/l\sim 1$ m$^{-1}$, then
\begin{eqnarray}
\sqrt{\langle\sigma_{33}(t)\sigma_{33}(t)\rangle^{(\zeta)}}
&\sim&\sqrt{\cosh(2\zeta)\left(\frac{\kappa^{2}}{30\pi^{2}t^{4}}\right)(0.3)+(\sinh2 \zeta )\left(\frac{\kappa^{2}}{120\pi^{2}t^{4}}\right)(-0.8)}\nonumber\\
&\sim&e^{\zeta}\sqrt{\langle\sigma_{33}(t)\sigma_{33}(t)\rangle^{(0)}}
\end{eqnarray}
Therefore, we obtain the enhancement factor $e^{\zeta}$ to the shear effect of the graviton noise due to the squeezed vacuum state as compare to that of the Minkowski vacuum state. 

\section{Conclusions and discussions}
To continue with our recent work \cite{ChoHu22}, we have considered in this paper the effect of quantum noise of gravitons as a stochastic tensorial force whose correlator is given by the graviton noise kernel associated with the Hadamard function of a quantized gravitational field. From the geodesic deviation equation between two masses, we see that this stochastic force has the stretching and compressing effects like the tidal force acting on neighboring particles. Therefore, one can view this as a stochastic tidal force. We can also extend the consideration to more masses, and calculate its influence on the evolution of a geodesic congruence.   Since this force is fluctuating, we estimate its magnitude by the square root of the equal time auto-correlator, which is proportional to the noise kernel. For the Minkowski vacuum, it is about 12 orders of magnitude smaller than the tidal force near the Earth's surface as shown in Eq.~(\ref{tidgra}). We have also considered the case with the squeezed vacuum state, and it is found that there is an enhancement of $e^{\zeta}$, where $\zeta$ is the squeeze parameter, over the Minkowski result. This enhancement factor may bring this stochastic tidal force to a detectable level someday.

With this stochastic tidal force, we have further investigated its effect on geodesic congruences. The kinematics of these congruences are characterized by the deformation tensor, including the expansion scalar, the shear tensor and the rotation tensor. The evolution of the deformation tensor is given by the Raychaudhuri equation in the presence of an external force. The fluctuations of the deformation tensor are then related to the graviton noise kernel. Due to the transverse-traceless properties of this noise kernel, the stochastic tidal force will not alter the fluctuations of the expansion scalar and the rotation tensor. Only the shear tensor which is symmetric and traceless will be affected by it. Again, this effect is too small to be detectable in the Minkowski vacuum case, but the enhancement factor of $e^{\zeta}$ in the squeezed vacuum state may increase the possibility. 

In the present work, we have solved the Raychaudhuri equations to the lowest order in the tidal force. In a more complete analysis, it is possible to include higher order effects in a perturbative manner. Although in the lowest order, the expansion scalar and the rotation tensor are not affected by the stochastic tidal force, we would expect them to be influenced by higher order terms. Another natural direction to build on our present analysis is to consider gravitons and their corresponding noise kernels in a curved spacetime,  going beyond the Minkowski background. For example, to examine the stochastic tidal force near a black hole in a consistent way, it is necessary to consider gravitons as quantized gravitational perturbations in a black hole background. Then we need to evaluate the noise kernel corresponding to these gravitons to study their influences on point masses. Likewise, evaluating the noise kernel in cosmological spacetimes is needed for the study of graviton noise effects in the universe.  We know that any quantum field in an expanding universe will be squeezed. To take advantage of the large squeezing factor for the enhancement of observable graviton noise effects one may need to trace back to quantum gravitational effects in the early universe, such as quantum fluctuations in primordial gravitons created and in matter structures formed. 

%
%
\acknowledgments
\noindent{HTC is supported in part by the Ministry of Science and
Technology, Taiwan, ROC, under the Grants MOST109-2112-M-032-007 and
MOST110-2112-M-032-009.  BLH thanks Prof. C. S. Chu for his hospitality during his visit to the National Center for Theoretical Sciences of Taiwan, ROC, when this work was completed. }


%
%



\begin{thebibliography}{99}


\bibitem{Feynman} Feynman, R., Acta Physica Polonica, Vol XXIV, p. 697 (1963)   compiled as  Feynman Lectures On Gravitation (1st ed.). CRC Press 2003. https://doi.org/10.1201/9780429502859 and  Feynman, Richard P., Fernando B. Morinigo, William G. Wagner, Brian Hatfield, John Preskill, and Kip S. Thorne. Feynman lectures on gravitation. CRC Press, 2018.

\bibitem{Weinberg} Weinberg, Steven. "Photons and gravitons in perturbation theory: Derivation of Maxwell's and Einstein's equations." Physical Review 138, no. 4B (1965): B988.

\bibitem{Oriti} D. Oriti (ed) Approaches to Quantum Gravity:
Toward a New Understanding of Space, Time and Matter,
(Cambridge University Press, Cambridge, England, 2009).

\bibitem{QGpheno} G. Amelino-Camelia, Quantum-spacetime phenomenology,
Living Rev. Relativity 16, 5 (2013).

\bibitem{QGexpt}  Experimental Search for Quantum Gravity, edited by S.
Hossenfelder (Springer, New York, 2018).

\bibitem{gravitonGW}
Arun, K. G., and Clifford M. Will. "Bounding the mass of the graviton with gravitational waves: effect of higher harmonics in gravitational waveform templates." Classical and Quantum Gravity 26, no. 15 (2009): 155002.

de Rham, Claudia, J. Tate Deskins, Andrew J. Tolley, and Shuang-Yong Zhou. "Graviton mass bounds." Reviews of Modern Physics 89, no. 2 (2017): 025004.

\bibitem{Don}
Donoghue, John F. "Leading quantum correction to the Newtonian potential." Physical Review Letters 72, no. 19 (1994): 2996.

\bibitem{HamLiu} Hamber, H. W., and S. Liu. "On the quantum corrections to the Newtonian potential." Physics Letters B 357, no. 1-2 (1995): 51-56.

\bibitem{DalMazNewton}
D. A. R. Dalvit and F. D. Mazzitelli, "Geodesics, gravitons, and the gauge fixing problem", Phys. Rev. D 56, 7779 (1997).

\bibitem{DalMazCarmen} Dalvit, Diego AR, Francisco D. Mazzitelli, and Carmen Molina-Paris. "One-loop graviton corrections to Maxwell’s equations." Physical Review D 63, no. 8 (2001): 084023.

\bibitem{Dyson} F. Dyson, Is a graviton detectable? in Poincare Prize
Lecture Presented at the International Congress of Mathematical
Physics, (World Scientific, Aalborg, Denmark, 2012).

\bibitem{QGTable} D. Carney, P. C. Stamp, and J. M. Taylor, Tabletop experiments
for quantum gravity: A user’s manual, Classical Quantum Gravity 36, 034001 (2019).

\bibitem{pertQG} F. Coradeschi, A. M. Frassino, T. Guerreiro, J. R. West, and
E. J. Schioppa, Can we detect the quantum nature of weak
gravitational fields? Universe 7, 414 (2021).


\bibitem{PWZ20} M. Parikh, F. Wilczek, G. Zahariade, {\em The noise of gravitons}. International Journal of Modern Physics D, 29(14), p.2042001 (2020).

\bibitem{PWZ21a} M. Parikh, F. Wilczek, G. Zahariade, {\em Quantum mechanics of gravitational waves}, Phys. Rev. Lett. 127, 081602 (2021).

\bibitem{PWZ21b}
M. Parikh, F. Wilczek, G. Zahariade, {\em Signature of the quantization of gravity at of gravitational wave detectors}, Phys. Rev. D 104, 046021 (2021).

\bibitem{Kanno} S. Kanno, J. Soda, and J. Tokuda, Noise and decoherence
induced by gravitons, Phys. Rev. D 103, 044017 (2021).


\bibitem{Haba} Z. Haba, State-dependent graviton noise in the equation of geodesic deviation, Eur. Phys. J. C 81 (2021) 40, [2009.12306].
Haba, Z. "Graviton noise: the Heisenberg picture." arXiv preprint arXiv:2202.06125 (2022).

\bibitem{ChoHu22}
Hing-Tong Cho and Bei-Lok Hu, ``Quantum noise of gravitons and stochastic force on geodesic separation'', Phys. Rev. D {\bf 105}, 086004 (2022).

\bibitem{FeyVer}  R. Feynman and F. Vernon, The theory of a general quantum
system interacting with a linear dissipative system, Ann.
Phys. (N.Y.) 24, 118 (1963); R. Feynman and A. Hibbs,
Quantum Mechanics and Path Integrals (McGraw-Hill,
New York, 1965).

\bibitem{Schwinger} J. Schwinger, Brownian motion of a quantum oscillator,
J. Math. Phys. (N.Y.) 2, 407 (1961).

\bibitem{Keldysh} L. V. Keldysh, Diagram technique for nonequilbrium processes,
Zh. Eksp. Teor. Fiz. 47, 1515 (1964) [Sov. Phys. JETP
20, 1018 (1965)].

\bibitem{Chou}  K. Chou, Z. Su, B. Hao, and L. Yu, Equilibrium and
nonequilibrium formalisms made unified, Phys. Rep. 118, 1 (1985).

\bibitem{CalHu08} E. Calzetta and B. L. Hu, Nonequilibrium Quantum Field
Theory (Cambridge University Press, Cambridge, England,
2008).

\bibitem{HuVer20} B. L. Hu and E. Verdaguer, Semiclassical and Stochastic
Gravity: Quantum Field Effects on Curved Spacetime
(Cambridge University Press, Cambridge, England, 2020).

\bibitem{CalLeg} A. O. Caldeira and A. J. Leggett, Path integral approach to
quantum Brownian motion, Physica (Amsterdam) A121,
587 (1983).

\bibitem{HPZ}  B. L. Hu, J. P. Paz, and Y. Zhang, Quantum Brownian
motion in a general environment: Exact master equation
with nonlocal dissipation and colored noise, Phys. Rev. D
45, 2843 (1992); Quantum Brownian motion in a general
environment. II. Nonlinear coupling and perturbative approach,
Phys. Rev. D 47, 1576 (1993).



\bibitem{RayEq} A. K. Raychaudhuri, Relativistic cosmology, I, Phys. Rev. 98, 1123 (1955).

\bibitem{Ellis} Ellis, G.F., On the Raychaudhuri equation. Pramana, 69(1), pp.15-22 (2007).

\bibitem{Kar} Kar, S. and Sengupta, S., The Raychaudhuri equations: A brief review. Pramana, 69(1), pp.49-76 (2007).

\bibitem{HawEll} S W Hawking and G F R Ellis, The large scale structure of space-time (Cambridge University Press, Cambridge, 1973)

\bibitem{RaySing} A K Raychaudhuri, A little reminiscence in singularities, black holes and cosmic censorship edited by P S Joshi (IUCAA, 1996). 
A K Raychaudhuri, A fresh look at the singularity problem, in: The Universe edited by N Dadhich and A K Kembhavi (Kluwer, Amsterdam, 2000).

\bibitem{QRayEqDas} Das, S., 2014, "Quantum Raychaudhuri equation", Physical Review D, 89(8), p.084068.  


\bibitem{RayEqZPL} Chakraborty, S., Kothawala, D. and Pesci, A., 2019. "Raychaudhuri equation with zero point length", Physics Letters B, 797, p.134877.

\bibitem{RayEqBGR} Burger, Daniel J., Nathan Moynihan, Saurya Das, S. Shajidul Haque, and Bret Underwood. "Towards the Raychaudhuri equation beyond general relativity." Physical Review D 98, no. 2 (2018): 024006.


\bibitem{QRayEqGUP} Vagenas, Elias C., Lina Alasfar, Salwa M. Alsaleh, and Ahmed Farag Ali. "The GUP and quantum Raychaudhuri equation." Nuclear Physics B 931 (2018): 72-78.

\bibitem{RayEqEffGUP} Blanchette, Keagan, Saurya Das, and Saeed Rastgoo. "Effective GUP-modified Raychaudhuri equation and black hole singularity: four models." Journal of High Energy Physics 2021, no. 9 (2021): 1-23.

\bibitem{RayEqLQG}
Blanchette, Keagan, Saurya Das, Samantha Hergott, and Saeed Rastgoo. "Black hole singularity resolution via the modified Raychaudhuri equation in loop quantum gravity." Physical Review D 103, no. 8 (2021): 084038.

\bibitem{RayEqEff} Ahmadi, N., and M. Nouri-Zonoz. "Quantum gravitational optics: Effective Raychaudhuri equation." Physical Review D 74, no. 4 (2006): 044034.

\bibitem{Ford}  Borgman J and Ford L H 2004, "The effects of stress tensor fluctuations upon focusing", Phys. Rev. D 70 064032. R T Thompson and L H Ford 2008, "Enhanced geometry fluctuations in Minkowski and black hole spacetimes", Class. Quantum Grav. 25 154006

\bibitem{Bak} Bak, Sang-Eon, Maulik Parikh, Sudipta Sarkar, and Francesco Setti. "Quantum Gravity Fluctuations in the Timelike Raychaudhuri Equation." arXiv preprint arXiv:2212.14010 (2022).

\bibitem{MTW}
C. W. Misner, K. S. Thorne, and J. A. Wheeler, "Gravitation", Princeton University Press (2017).

\bibitem{Shore18}
G. M. Shore, "Memory, Penrose limits and the geometry of gravitational shockwaves and gyratons", JHEP 12, 133 (2018).

\bibitem{Poisson}
E. Poisson, ''A relativist's toolkit: The mathematics of black-hole mechanics", Cambridge University Press (2004).



\end{thebibliography}
\end{document}